\documentclass[graphics,floatfix, footinbib,tightenlines,nobibnotes, aps, prl, twocolumn]{revtex4-1}
\usepackage{amsmath,amssymb,wasysym}
\usepackage{graphicx}
\usepackage{dcolumn}
\usepackage{bm}
\usepackage{braket}
\usepackage{verbatim}
\usepackage{float}
\usepackage{bbold}
\usepackage{color}
\usepackage{xcolor}
\usepackage{relsize}
\usepackage{amsthm}
\usepackage{enumerate}
\usepackage{soul,xcolor}
\usepackage[normalem]{ulem} 

\usepackage[T1]{fontenc}
\usepackage[colorlinks=true ,urlcolor=blue,urlbordercolor={0 1 1}]{hyperref}

\newcommand{\beq}{\begin{eqnarray}}
\newcommand{\eeq}{\end{eqnarray}}

\newcommand{\bsp}{\begin{split}}
\newcommand{\esp}{\end{split}}

\newcommand{\be}{\begin{equation}}
\newcommand{\ee}{\end{equation}}

\newcommand{\moire} {moir{\' e} }

\begin{document}

\setstcolor{red}
\title{Ferromagnetism in narrow bands of moir\'e superlattices}
\author{C\'ecile Repellin}
\author{Zhihuan Dong}
\author{Ya-Hui Zhang}
\author{T. Senthil}
\affiliation{Department of Physics, Massachusetts Institute of Technology, Cambridge, MA, USA
}

\date{\today}

\begin{abstract}
Many graphene \moire superlattices host narrow bands with non-zero valley Chern numbers. We provide analytical and numerical evidence for a robust spin and/or valley polarized insulator at total integer band filling in nearly flat bands of several different \moire materials. In the limit of a perfectly flat band, we present analytical arguments in favor of the ferromagnetic state substantiated by numerical calculations. Further, we numerically evaluate its stability for a finite bandwidth. We provide exact diagonalization results for models appropriate for ABC trilayer graphene aligned with hBN, twisted double bilayer graphene, and twisted bilayer graphene aligned with hBN. We also provide DMRG results for a honeycomb lattice with a quasi-flat band and non-zero Chern number, which extend our results to larger system sizes.
We find a maximally spin and valley polarized insulator at all integer fillings when the band is sufficiently flat. We also show that interactions may induce effective dispersive terms strong enough to destabilize this state. These results still hold in the case of zero valley Chern number (for example, trivial side of TLG/hBN). We give an intuitive picture based on extended Wannier orbitals, and emphasize the role of the quantum geometry of the band, whose microscopic details may enhance or weaken ferromagnetism in moir\'e materials. 
\end{abstract}

\maketitle

Following the remarkable discovery of correlated insulators and superconductivity in Magic Angle Twisted Bilayer Graphene (TBG)~\cite{cao2018correlated,cao2018unconventional}, a great deal of attention has been lavished on various ``\moire materials". In TBG, the correlated insulator/superconductor has since been observed by other groups~\cite{yankowitz2019tuning,lu2019superconductors}, leading to a wealth of new information. When the TBG is further aligned with a hexagonal Boron Nitride substrate (TBG/hBN), emergent ferromagnetism and a large~\cite{Aaron2019Emergent} or even quantized~\cite{young2019QAH} anomalous Hall effect is observed at $3/4$ filling of the conduction band. 

In other experiments, the \moire bands formed when ABC stacked trilayer graphene is aligned with one of the hBN (TLG/hBN) substrates display interesting strong correlation physics\cite{Wang2019Evidence,Wang2019Signatures,chen2019tunable}. Applying a perpendicular displacement field $D$ enables control of the bandwidth leading to a gate-tunable correlated insulator at half-filling\cite{Wang2019Evidence}. Furthermore the sign of $D$ enables changing the band topology\cite{zhang2019nearly,chittari2019gate,chen2019tunable}: for one sign of $D$ the bands in each of the two valleys have equal and opposite Chern number while for the other sign of $D$, the Chern number in either valley is zero. Remarkably in the topologically non-trivial side, the $1/4$ filled state in the valence band shows ferromagnetism and a quantized anomalous Hall effect with Chern number 2\cite{chen2019tunable}. In yet other experiments, in twisted double bilayer graphene systems (TDBG, {\em i.e} bilayer graphene twisted relative to another bilayer graphene to a magic angle), good evidence is found for a spin polarized insulator~\cite{Liu2019Spin,Shen2019Observation,Cao2019Electric} at half-filling of the conduction band which gives way~\cite{Liu2019Spin,Shen2019Observation}, upon doping, to a superconductor which is likely also spin polarized. Theoretically, the conduction band in the TDBG system also has a non-zero Chern number which is equal and opposite for the two valleys~\cite{zhang2019nearly,lee2019theory,koshino2019band,liu2019quantum}.

In this paper, we study the physics of narrow \moire bands in the strong interaction limit, {\em i.e} when the Coulomb interaction is much larger than the bandwidth. Our focus is on the variety of \moire systems where the conduction and valence bands are separated from each other by energy gaps~\footnote{This is however not the case  in the original TBG where there are Dirac points connecting conduction and valence bands, and our results do not directly apply to this system}. 
We call $\nu_T$ the filling fraction including the spin and valley degrees of freedom, {\em i.e} the number of electrons per \moire unit cell in the active band.
We restrict our attention to total integer fillings $\nu_T = 1, 2, 3$, and study the nature of the correlated insulators in this strong interaction limit. Previous papers have presented physical arguments, and supporting Hartree-Fock calculations, that the natural fate of the system in this limit is a spin-valley ferromagnetic insulator~\cite{zhang2019nearly,lee2019theory,bultinck2019anomalous, zhang2019bridging, zhang2019twisted, xie2018nature}. However Hartree-Fock theory typically overestimates the stability of ferromagnetic states. Thus it is important to substantiate the physical arguments for spin-valley ferromagnetism through other less biased calculations. 
In the context of strained graphene, Ref.~\onlinecite{ghaemi-PhysRevLett.108.266801} provided numerical evidence for a valley-polarized insulator at fractional filling.
Here we present analytical arguments and numerical calculations - both exact diagonalization (ED) and density matrix renormalization group (DMRG) -  for strongly interacting nearly flat band models pertinent to many \moire materials. We show that indeed spin-valley ferromagnetic states are stabilized in the flat-band limit. Our results highlight the robustness of ferromagnetism in a narrow band even when it is topologically trivial~\cite{zhang2019bridging}. Moreover, we provide a quantitative estimate of the stability range of the ferromagnetic states in terms of the interaction to bandwidth ratio. We also show that intervalley coherent order is always disfavored compared to ferromagnetism in the limit where the flat band is a Landau level.
We emphasize that the interactions renormalize the bare (non-interacting) dispersion of the band through various mechanisms detailed in the paper. The strong interaction limit is thus defined as the regime where residual interactions far exceed the renormalized bandwidth.

We consider a two dimensional material with spinful electrons occupying bands in each of two disconnected valleys (denoted $+$ and $-$) such that time-reversal maps one valley to the other. As a result, the $+$ and $-$ valleys of the active band have equal and opposite Chern number $C$. We consider density-density interactions of typical strength $U$, and focus on the strong coupling limit $U/W \gg 1$, where $W$ is the bandwidth of the active band.

\textit{Physical argument for ferromagnetism}
First consider the case of a topologically {\em trivial} band with non-zero Berry flux distribution. Such a situation arises, in TLG/h-BN for one sign of the displacement field $D$~\cite{zhang2019nearly}. Using a Wannier basis~\cite{brouder2007wannier,marzari2012maximally}, we can build an effective tight-binding model for the active band~~\cite{zhang2019bridging}. The Wannier functions have a finite extension which is needed to capture the Berry flux density in the band. Projecting the Coulomb interaction onto the Wannier basis leads to an on-site Hubbard interaction of order $U$ (as well as smaller terms between neighboring sites) but also to an inter-site {\em ferromagnetic} Hund's interaction $J_F = g_s U$~\cite{zhang2019bridging}. The coefficient $g_s$ depends on the overlap of Wannier densities at neighboring sites but stays finite even when $W \rightarrow 0$. In the large-$U$ limit, the ground state has a fixed number $\nu_T \in \mathbb Z$ of electrons at each site. The active degrees of freedom are local moments in the spin-valley space. In the strict limit $W \rightarrow 0$, the only coupling that survives between these local spin-valley degrees of freedom is $J_F$, giving rise to a spin-valley ferromagnet~\cite{zhang2019bridging} (see also Ref.~\onlinecite{KangVafek-PhysRevLett.122.246401}). As $W$ increases, there will also be antiferromagnetic inter-site superexchange $\frac{\alpha W^2}{U}$ with $\alpha$ a constant independent of $W$ and $U$. This antiferromagnetic exchange can dominate over the inter-site Hund's exchange only when $W > \sqrt{\frac{g_s}{\alpha}}U$.  Thus so long as $W/U$ is small enough we get a spin-valley ferromagnetic Mott insulator. 
In this mechanism the larger the extension of the Wannier functions, the larger the coefficient $g_s$, and thus the stronger the ferromagnetism.

Turning next to the case of topologically non-trivial $\pm $ Chern bands, symmetric Wannier functions cannot be localized~\cite{brouder2007wannier,marzari2012maximally}; we may view this as the limit of Wannier functions with infinite extension. Intuitively, the Hund's effect will only be stronger than in the topologically trivial case, and hence a spin-valley ferromagnetic insulator is the likely ground state at all integer fillings.

\textit{Analytical considerations}
For a perfectly flat band separated by a large gap from other bands, the effective Hamiltonian takes the form of the Coulomb interaction projected onto the active band
\begin{equation}
H_V  = \sum_{\mathbf q} \delta \tilde{\rho}(\mathbf q) V(\mathbf q)\delta \tilde{\rho}(-\mathbf{q})
\label{eq:Hamiltonian_projection}
\end{equation}
where $\delta \tilde{\rho} (\mathbf{q}) = \tilde{\rho} (\mathbf{q}) - \rho_0 \delta^2(\mathbf{q)}$ is the deviation of the projected density from the average density $\rho_0$.
The total projected density operator $\tilde{\rho}$ is summed over spin and valley indices and can be written
\begin{eqnarray}
\tilde \rho(\mathbf q) & = & \tilde{\rho}_+(\mathbf q) + \tilde{\rho}_-(\mathbf q) \\
\tilde{\rho}_s(\mathbf q) & = &  \sum_{{\mathbf k}, \sigma} \lambda_s (\mathbf k + \mathbf q, \mathbf k) c^\dagger_{\mathbf k + \mathbf q, \sigma s} c_{\mathbf k, \sigma s}
\end{eqnarray}
Here $\tilde{\rho}_{\pm}$ are the projected densities in each valley $s$. $\sigma = \uparrow, \downarrow$ is the spin index. The $\lambda_s$ are valley dependent form factors which are defined in terms of the Bloch eigenstates $|u_{s,\mathbf k} \rangle$ through
\begin{equation}
 \lambda_s (\mathbf k + \mathbf q, \mathbf k) = \langle u_{s, \mathbf k + \mathbf q}|u_{s,\mathbf k} \rangle
\end{equation}
Because of the Berry flux distribution, $\lambda_s$ is a non-trivial function of $\mathbf k$ and $\mathbf q$.
It  is readily verified that
$\tilde{\rho}(\mathbf{-q}) = \tilde{\rho}(\mathbf{q})^\dagger$ 
 as befits the total density operator. The Hamiltonian $H_V$ is  invariant under a $U(2) \times U(2)$ rotation corresponding to independent charge and spin conservation within each valley. It is {\em not} $SU(4)$ invariant in spin-valley space due to the form factors. With the further assumption that the interaction $V(\mathbf{q}) \geq 0$ for all $\mathbf{q}$ (satisfied for Coulomb and for short range repulsive potentials), it becomes clear that $H_V$ is positive semi-definite. Thus any state $\ket{\psi}$ that satisfies 
\begin{equation}
\delta \tilde{\rho}(\mathbf{q}) |\psi \rangle = 0
\label{GS_condition}
\end{equation}
for all $\mathbf q$ is an exact ground state. 

At $\nu_T$ integer, consider the state obtained by filling up $\nu_T$ bands to form an insulator. At $\nu_T = 2$ this can be spin-polarized or valley-polarized. At $\nu_T = 1, 3$ this must be both spin and valley polarized. These states satisfy Eq.~\ref{GS_condition}, hence they are exact eigenstates of $H_V$ with eigenvalue $0$. It follows that any ground state of $H_V$ must satisfy Eq.~\eqref{GS_condition} for all $\mathbf q$. The remaining question is whether the spin-valley polarized states are the unique ground states. Indeed the same argument applied to a half-filled Hubbard model in the flat band limit (zero hopping) would yield ferromagnetic ground states; but these are degenerate with all other spin configurations and hence are not unique.  In contrast in the flat Chern band, spin flips in the ferromagnetic state generically cost energy. This is well known in the  idealized case of a band with uniform Berry curvature, {\em i.e} quantum Hall ferromagnetism~\cite{sondhi1993skyrmions, Ezawa_2009}. In this case, the spin stiffness $\rho_s$ can be calculated exactly~\cite{zhang2019nearly} and is proportional to the square of the Chern number. This suggests that these ferromagnetic states may indeed be the unique ground states for any flat Chern band.

Besides spin-valley polarized states, inter-valley coherent (IVC) states are a plausible ground state candidate for \moire systems at integer filling $\nu_T$~\cite{jung2015origin,bultinck2019anomalous,zhang2019twisted, zhang2019nearly, lee2019theory} (see also Ref.~\onlinecite{po2018origin}). To address these, it is instructive to consider a toy-model where the active band is the lowest Landau level of a system with opposite magnetic fields $\pm B$ for each valley. Then the projected density operators satisfy the Girvin-MacDonald-Platzman (GMP) algebra~\cite{GMP1986} ($l_B$ is the magnetic length): 
 \begin{equation}
 [\delta \tilde{\rho}_{\pm}(\mathbf q), \delta \tilde{\rho}_{\pm}(\mathbf q')\ = \pm 2i \sin\left(\frac{(\mathbf q \times \mathbf q')l_B^2 }{2}\right)\delta \tilde{\rho}_{\pm}(\mathbf q + \mathbf q')
  \end{equation}

 Since the commutator has the opposite sign for the two opposite valleys, the total density satisfies 
  \begin{equation}
  \label{rhocmm}
  [\delta \tilde{\rho}(\mathbf q), \delta \tilde{\rho}(\mathbf q')] =  2i \sin\left(\frac{(\mathbf q \times \mathbf q')l_B^2 }{2}\right)I^z(\mathbf q + \mathbf q')
  \end{equation}
  where $I^z(\mathbf q) = \delta \tilde{\rho} _+(\mathbf{q}) - \delta \tilde{\rho}_-(\mathbf{q})$ is the Fourier transform of the valley charge density. Thus in any ground state, by applying Eq.~\eqref{GS_condition} to the left hand side of Eq.~\eqref{rhocmm}, we find that 
  \begin{equation}
  \label{zeroIz}
  I^z(\mathbf q) |\psi \rangle = 0,\ \forall \mathbf q \neq 0
  \end{equation}
Thus the valley charge density cannot have fluctuations at any non-zero $\mathbf q$ in a ground state. If however there is IVC order then the breaking of valley $U_v(1)$ symmetry will  lead to non-zero fluctuations of $I^z(\mathbf q)$. For instance from the expected Goldstone fluctuations of the phase of the IVC order parameter the ground state correlator  
  \begin{equation}
  \langle \psi| I^z(-\mathbf q) I^z(\mathbf q) |\psi \rangle \underset{\mathbf q \rightarrow 0}{=} \frac{\sqrt{\kappa_v \rho_{sv}}}{2} |\mathbf q|
  \end{equation} 
(Here $\kappa_v$ is the valley charge susceptibility,and $\rho_{sv}$ is the phase stiffness of the IVC order parameter). The exact result in Eq.~\eqref{zeroIz} is clearly in conflict with this expectation. 
We conclude therefore that the IVC ordered state is not a ground state in this model. 
Moving away from the toy model above, introducing a small bandwidth and Berry curvature fluctuations will clearly not change this result~\footnote{In a generic Chern band, the GMP algebra is indeed approximated in the $\mathbf q\rightarrow 0$ limit~\cite{ParameswaranFCI2013, PhysRevB.90.045114}, and Eq.~\eqref{zeroIz} still holds in this limit}. We thus expect that, as suggested by Hartree-Fock, the IVC state is disfavored relative to the spin-valley polarized states in the nearly flat band limit for generic $\pm C$ Chern bands. 
In what follows we will provide numerical results supporting this expectation, and find that spin-valley ferromagnetism is indeed the ground state of nearly flat $\pm$ Chern bands at integer total filling. 

Note that in realistic models, various effects are responsible for the dispersion of the active band. Besides the bare bandwidth $W_\mathrm{bare}$ of the non-interacting model, two bilinear terms of order $U$ contribute to the overall bandwidth. The first term is the Fock term stemming from the interaction between electrons in the active band and in the lower (fully occupied) bands. The second bilinear term is the difference between $H_V$ and its normal-ordered couterpart, and is proportional to the fluctuations of the squared form factor $|\lambda_s (\mathbf k + \mathbf q, \mathbf k)|^2$. See the supplementary material for more details. In our analytical considerations, we considered the ideal case where the \emph{renormalized} bandwidth vanishes.

\begin{figure}
\centering
\includegraphics[width = 0.49\textwidth]{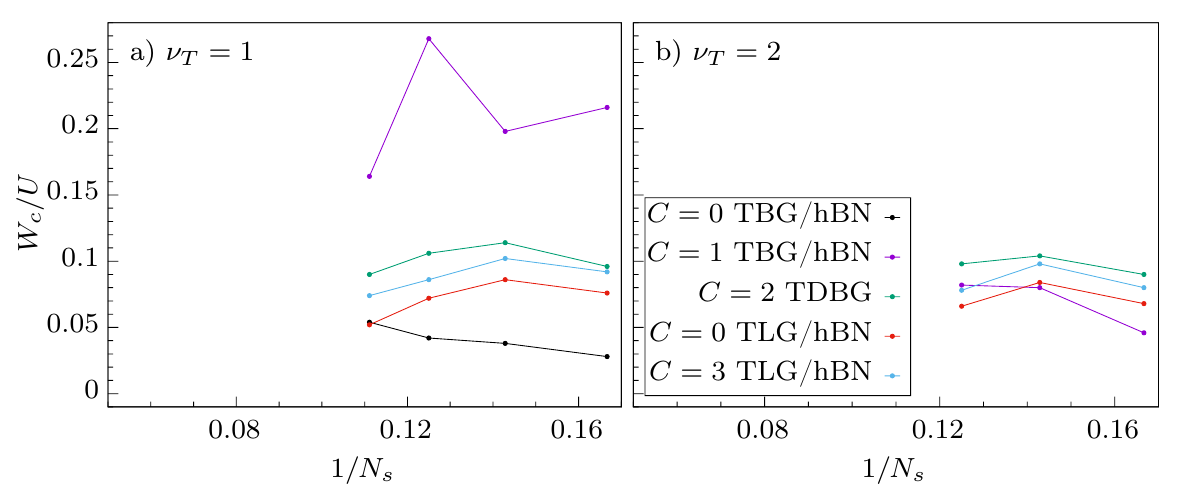}
\caption{Stability of the FM state in the isolated flat band upon adding a finite bandwidth $W_\mathrm{bare}$. The maximum value of bandwidth $W_c$ is plotted as a function of the number of moir\'e unit cells $N_s$ at filling $\nu_T=1$ and $\nu_T = 2$. Note that for the same number of moir\'e unit cells $N_s$, the dimension of the Hilbert space is much larger at $\nu_T = 2$ than at $\nu_T = 1$.}
\label{fig: critical bandwidth ED}
\end{figure}

 \textit{Exact diagonalization results for a single \moire band}
 We consider the continuum momentum-space models~\cite{bistritzer2011moire} of three iconic \moire systems (TBG/hBN~\cite{zhang2019twisted}, TDBG~\cite{zhang2019nearly,lee2019theory}, TLG/hBN~\cite{zhang2019nearly}). 
 Unless otherwise noted, for concreteness we respectively choose the twist angles $\theta = 1.05^{\circ}$, $1.2^{\circ}$ and $0$, and the displacement field $D = 0$, $40 \mathrm{mV}$ and $50 \mathrm{mV}$. The active band is the valence (TLG/hBN) or the conduction (TBG/hBN and TDBG) band and has Chern number $C=\pm1$ (TBG/hBN), $C=\pm2$ (TDBG) and $C=\pm3$ (TLG/hBN)\footnote{In TLG/hBN, a quantized anomalous Hall effect was in fact observed at $C=2$~\cite{chen2019tunable}. Here, we use the continuum non-interacting model which predicts a Chern $C=\pm3$ band. A scenario involving interactions and supported by Hartree-Fock calculations was proposed in Ref.~\onlinecite{chen2019tunable} to explain this discrepancy.}. We also consider the trivial ($C=0$) band obtained by switching the sign of the coupling to the top hBN layer in TBG/hBN or the sign of the displacement field $D$ in TLG/hBN. 
 We take the limit where the active band is separated from other bands by a gap much larger than its bandwidth.
 The Hamiltonian is obtained by normal-ordering the projected Hamiltonian $H_V$ of Eq.~\eqref{eq:Hamiltonian_projection}, where the screened Coulomb interaction takes the form
 \begin{equation}
 V(q) = U\frac{1}{q}\left(1 - e^{-qr_0} \right)
 \label{eq: screened coulomb}
 \end{equation}
 $r_0$ is the screening length (we choose $r_0= 5.0$ in units of the \moire lattice constant). We also consider the addition of a kinetic term which gives the active band a width $W$.
\begin{equation}
H = H_V - \frac{W_\mathrm{bare}}{2}\sum_{\mathbf{k}}\cos\left( \mathbf{k}\cdot\mathbf{a_1} + \mathbf{k}\cdot\mathbf{a_2}\right) c^\dagger_{\mathbf{k}}c_{\mathbf{k}}
\end{equation}
where $\mathbf{a_1}, \mathbf{a_2}$ are the \moire lattice vectors, and we have taken a simplified dispersion (not the realistic one). We study this model using exact diagonalization at integer filling $\nu_T = 1, 2$ ($\nu_T = 3$ is related to $\nu_T=1$ in our model through a particle-hole transformation). We call $N_s$ the number of moir\'e unit cells, and choose the aspect ratio of the finite cluster to be close to $1$.
 
We start by investigating the nature of the ground state in the limit $W_\mathrm{bare}=0$. In spite of the band flatness, the normal ordering of the interaction induces a finite dispersion (see supplementary materials), such that ferromagnetism is not guaranteed. Nevertheless, for $\nu_T=1$, we find that the ground state is always fully spin and valley polarized; the resulting state has a quantum anomalous Hall effect $\sigma_{xy} = Ce^2/h$. At $\nu_T=2$, maximal polarization of spin or valley leads to several correlated insulators all related by $U(2)\times U(2)$ symmetry in our model. For example, one with full spin polarization (but $I_z = 0$), which is a valley-Hall insulator if $C\neq 0$, and one with full valley polarization ($S_z = 0$), with anomalous Hall effect $\sigma_{xy} = 2C e^2/h$.\footnote{Including the weak inter-valley Hund's interaction would break the $U(2) \times U(2)$ symmetry and lift the degeneracy between these two states}. Numerically, we find that these $\nu_T=2$ polarized insulators indeed have the lowest energy in all models except for one: the $C=0$ TBG/hBN. In this case, the ground state is partially polarized, but the important finite-size effects prevent us from identifying its nature.

Adding a finite bandwidth $W$, we find that the ferromagnetic phase survives up to $W_c \simeq 0.05U$ in some models, and up to $W_c = 0.2U$ in others at filling $\nu_T = 1$. At $\nu_T=2$, the ferromagnet is relatively less stable, with critical values ranging from $W_c = 0.04U$ to $W_c = 0.1U$. The critical bandwidth is extracted from a finite-size extrapolation of exact diagonalization results which is detailed in Fig.~\ref{fig: critical bandwidth ED}.

\begin{figure}
\centering
\includegraphics[width = 0.49\textwidth]{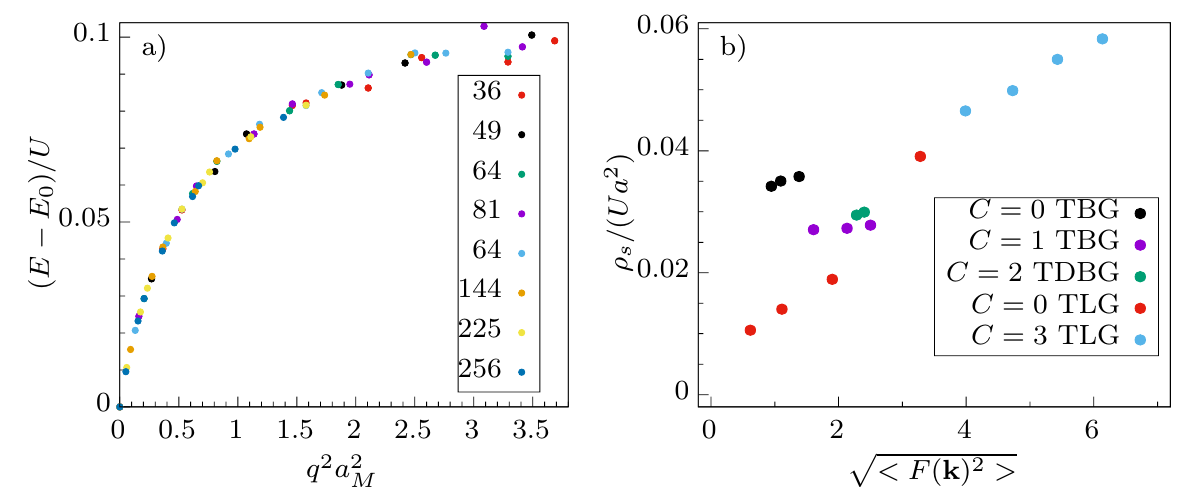}
\caption{Properties of low-energy spin excitations at $\nu_T=1$ and $W_\mathrm{bare} = 0$. a) Spin-wave dispersion around the $\Gamma$ point in the $C=\pm 3$ model from exact diagonalization. The colors correspond to different numbers $N_s$ of moir\'e unit cells. b) Spin stiffness $\rho_s$ ($a$ is the moir\'e lattice constant), as a function of the squared Berry curvature $F(\mathbf k)^2$ averaged over the Brillouin zone. We evaluated $\rho_s$ from a linear fit of the spin-wave dispersion at the $\Gamma$ point. For each model, we changed the Berry curvature distribution by adjusting the twist angle $\theta$ or the displacement field $D$.}
\label{fig: spin stiffness}
\end{figure}

The spin stiffness measures the energy change from twisting the spin boundary conditions
\begin{equation}
\rho_s = \frac{\partial E(\theta_s)}{\partial \theta_s}\biggr\rvert_{\theta_s = 0}
\end{equation}
To evaluate $\rho_s$, we calculated the spin-wave dispersion exactly for large systems (hundreds of moir\'e unit cells) by restricting the calculation to the valley-polarized, $S_z = N/2 - 1$ sector. Fig.~\ref{fig: spin stiffness}a) shows this dispersion in the case of the $C=3$ model, and displays a remarkable data collapse for $36 < N_s < 256$. 
For these larger systems, calculating the spectrum of all spin and valley polarization sectors is impossible within exact diagonalization, but we rely on our previous result and assume that the other polarization sectors do not affect the low-energy properties. Fig.~\ref{fig: spin stiffness}b) illustrates the influence of the Berry curvature distribution $F(\mathbf k)$ on the spin stiffness ($F(\mathbf k)$ is changed by tuning the microscopic parameters such as the twist angle $\theta$ or the displacement field $D$). 
While $F(\mathbf k)$ does not uniquely determine the spin stiffness~\footnote{Even in the case of uniform Berry curvature, the real part of the quantum geometric tensor also enters the expression of $\rho_s$~\cite{zhang2019nearly}}, for a given model (a given color in Fig.~\ref{fig: spin stiffness}b), larger Berry curvature fluctuations appear to enhance $\rho_s$.

\begin{figure}
\centering
\includegraphics[width = 0.49\textwidth]{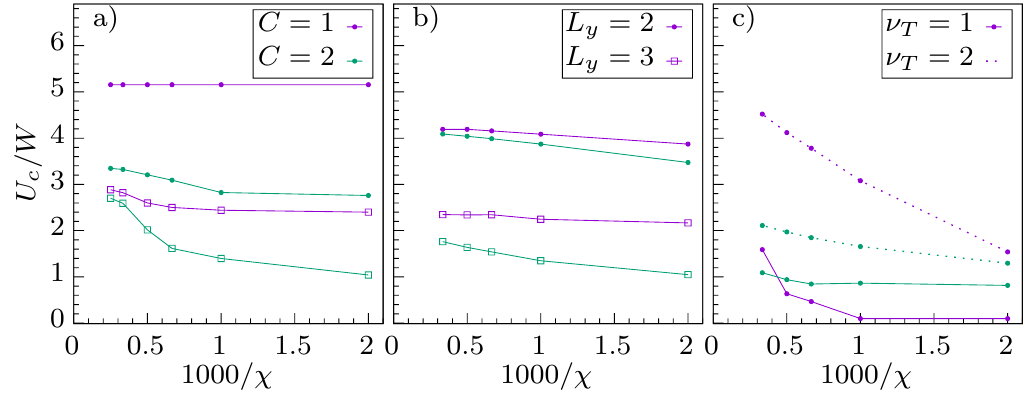}
\caption{Onset of ferromagnetism in the $C=1,2$ Honeycomb model Eq.~\eqref{eq: haldane}, extracted from DMRG on cylinders of perimeter $L_y$, as a function of the bond dimension $\chi$. a) Valley-polarized model, $\nu_T=1$. b) Spin-polarized model, $\nu_T=1$. c) Full spin and valley model at $\nu_T=1, 2$, limited to $L_y=2$. The difference of $U_c/W$ between a), b) and c) are not significant for these small system sizes. Note that the y axis ($U_c/W$) differs from the y axis of Fig.~\ref{fig: critical bandwidth ED} ($W_c/U$).}
\label{fig: Honeycomb Hubbard}
\end{figure}

\textit{Ferromagnetism in the spin-valley Haldane model}
We now turn to DMRG calculations on infinitely long cylinders, which help us circumvent the size limitations of exact diagonalization. We use a tight-binding model, which facilitates DMRG~\footnote{While it is in principle possible to perform DMRG using a hybrid momentum-real space basis~\cite{motruk-PhysRevB.93.155139, ehlers-PhysRevB.95.125125}, it comes with the cost of longer-range interactions and more complex algorithms.}, but comes with an additional cost: due to the non-trivial Chern number of the active band, we must consider a two-band model. Additionally, the two-band model permits the consideration of the band gap energy scale, which we have supposed to be infinite until now.
Our toy model is a tight-binding model on the honeycomb lattice based on the Haldane model~\cite{haldane1988} with on-site Hubbard interaction of strength $U$.
\begin{eqnarray}
H_1 & = & -\sum_{i, j, \sigma} \left(t_{ij} c^\dagger_{i\sigma+}c_{j\sigma+} + t_{ij}^* c^\dagger_{i\sigma-}c_{j\sigma-} + h.c. \right)\nonumber \\
    &   & + U \sum_{i, \sigma, \sigma', s, s'} n_{i\sigma s}n_{i\sigma' s'}
    \label{eq: haldane}
\end{eqnarray}
The hopping amplitudes $t_{ij}$ are non-zero for first ($t$) and second ($t_2$) neighbor and realize the Haldane model. We also consider third ($t_3$) and fourth ($t_4$) neighbor hopping. We tune these parameters to obtain a narrow conduction band with Chern number $C=\pm1$ and $C=\pm2$; we use the parameters $t_2/t=0.315e^{0.209\pi i}$ , $t_3 = t_4 =0$ for $C=\pm1$ and $t_2/t = 1.312i$, $t_3/t=1.312$, $t_4/t=0.524$.

We numerically obtained the ground state of $H_1$ using DMRG in the infinite cylinder geometry with $L_y = 2$ unit cells along the perimeter. We extracted the spin and valley polarization of the ground state for several values of $U$. In spite of a large bond dimension dependence, we find that the onset of ferromagnetism $U_c/W$ is always smaller than the band gap (see Fig.~\ref{fig: Honeycomb Hubbard}c). 
Working in the valley-polarized or spin-polarized limits, we can better approach convergence in our DMRG calculations, and simulate wider cylinders (up to $L_y=3$). Our results are shown in Fig.~\ref{fig: Honeycomb Hubbard} and confirm the ferromagnetic nature of the ground state for $U > U_c$ where $U_c \simeq 3W$
(the band gap is respectively $\Delta/W = 6.0$ and $5.45$ for the $C=\pm 1$ and $C=\pm 2$ parameters). $U_c$ seemingly decreases with increasing system size $L_y$ (our complementary exact diagonalization results on this model show the same trend, see supplementary material for details), giving us confidence that $U_c < \Delta$ in the thermodynamic limit.

\textit{Discussion}
In this paper we have shown both analytical and numerical evidence for spin and valley polarization in nearly flat bands, which naturally emerge in several graphene moir\'e superlattices. 
Our results demonstrate a valley and spin polarized quantum anomalous Hall insulator at $\nu_T=1$ or $\nu_T=3$. At $\nu_T=2$, a spin-polarized valley Hall insulator and a valley-polarized quantum anomalous Hall insulator are both possible. 
Indeed recent experiments have already observed signatures for spin polarization  in twisted double bilayer graphene and anomalous Hall effect in TBG/hBN\cite{Aaron2019Emergent} and TLG/hBN\cite{chen2019tunable}.

The phenomenon we described is reminiscent of quantum Hall ferromagnetism, but there are important differences. Most naively, the finite bandwidth may destroy ferromagnetism, and we have quantitatively evaluated the position of this transition. A less expected difference is that flat-band ferromagnetism appears even when the Chern number is zero, due to the non-zero Berry flux. 
Deviating from the Landau level situation through large Berry curvature fluctuations has two opposite effects, which respectively destabilize and stabilize ferromagnetism: it may increase the strength of interaction-induced dispersive terms, but it may also enhance the spin stiffness.
A natural future direction is to study the possibility of fractional quantum Hall effect from similar spontaneous time reversal breaking at fractional filling.

\textit{Acknowledgement}
We are thankful to Yin-Chen He and Kun Yang for insightful discussions.
DMRG calculations were performed using the TeNPy Library (version 0.4.0)\cite{tenpy}.
This work was supported by NSF grant DMR-1911666, and partially through a Simons Investigator Award from the Simons Foundation to Senthil Todadri.
C.R. is supported by the Marie Sklodowska-Curie program under EC Grant agreement No. 751859.
Part of this work was performed during a visit of CR and TS at Aspen Center for Physics, which is supported by National Science Foundation grant PHY-1607611.

\textit{Author contributions}
CR performed the ED simulations, ZD performed the DMRG simulations. YHZ provided the momentum space and lattice models. CR, YHZ and TS supervised the work of ZD. All authors contributed to the discussions and analysis of results.
\bibliographystyle{apsrev4-1}
\bibliography{qhfm}

\onecolumngrid

\section{Renormalization of the bare bandwidth by interactions}

In this section, we discuss the effective band broadening originating from the projection of the interaction onto the active band.

We first recall the description of the density interaction in the case of a periodic system without any band projection.
We define $f(\mathbf k)$ as operators for the microscopic electrons. $c(\mathbf k)$ is the band operator. We have the following relation
\begin{equation}
  f_{\alpha,I}(\mathbf k)=\sum_m  u_{\alpha,m}^I(\mathbf k)c_{\alpha,m}(\mathbf k)
  \label{eq:f_c_relation}
\end{equation}
where $\alpha$ is the spin-valley index, $m$ is the band index and $I$ is the orbital index at the microscopic level (for example, in the Haldane model $I=A,B$ is the sublattice index). $u_{\alpha,m}(\mathbf k)$ is the Bloch wavefunction of the band specified by $\alpha,m$. It is a vector for which $I$ is the index.  Therefore there is a completeness relation:
\begin{equation}
  \sum_{m}u^{I*}_{\alpha,m} u_{\alpha,m}^J=\delta_{IJ}
  \label{eq: completeness}
\end{equation}
and an orthogonality relation
\begin{equation}
  \sum_{I}u^{I*}_{\alpha,m} u_{\alpha,n}^I=\delta_{mn}
\end{equation}
In terms of $f$ the interaction term is written

\begin{equation}
  H_\mathrm{int}=\frac{1}{2}\sum_{\mathbf{k_1},\mathbf{k_2},\mathbf{q}}\sum_{\alpha,\beta}\sum_{I,J} V(\mathbf{q})f^\dagger_{\alpha,I}(\mathbf{k_1+q})f^\dagger_{\beta,J}(\mathbf{k_2-q})f_{\beta,J}(\mathbf{k_2})f_{\alpha,I}(\mathbf{k_1})
\end{equation}

Substituting Eq.~\ref{eq:f_c_relation}  we get
\begin{align}
  H_\mathrm{int}&=\frac{1}{2}\sum_{\mathbf{k_1},\mathbf{k_2},\mathbf{q}}\sum_{\alpha,\beta}\sum_{m_1,m_2}\sum_{n_1,n_2} V(\mathbf{q})c^\dagger_{\alpha,m_1}(\mathbf{k_1+q})c^\dagger_{\beta,n_1}(\mathbf{k_2-q})c_{\beta,n_2}(\mathbf{k_2})c_{\alpha,m_2}(\mathbf{k_1})\notag\\
  &\braket{u_{\alpha,m_1}(\mathbf{k_1+q})| u_{\alpha,m_2}(\mathbf{k_1}}\braket{u_{\beta,n_1}(\mathbf{k_2-q})|u_{\beta,n_2}(\mathbf{k_2})}
  \label{eq:normal_order_interaction}
\end{align}
We define the density operator
\begin{equation}
  \rho_\alpha(\mathbf q)=\sum_{m_1,m_2}\sum_{\mathbf k}c^\dagger_{\alpha,m_1}(\mathbf{k+q})c_{\alpha,m_2}(\mathbf{k})\braket{u_{\alpha,m_1}(\mathbf{k+q})| u_{\alpha,m_2}(\mathbf{k})}
\end{equation}
With simple algebra, we can write $H_\mathrm{int}$ in the form $\rho(\mathbf q)\rho(-\mathbf q)$
\begin{align}
  H_\mathrm{int}= & \frac{1}{2}\sum_{\mathbf{q}} \sum_{\alpha \beta} V(\mathbf{q}) \rho_\alpha(\mathbf q)\rho_\beta(-\mathbf q)\notag\\
  &-\frac{1}{2}\sum_{\mathbf{k},\mathbf{q}}\sum_{\alpha}\sum_{m_1,m_2}\sum_{n_2} V(\mathbf{q})c^\dagger_{\alpha,m_1}(\mathbf{k})c_{\alpha;n_2}(\mathbf{k})\braket{u_{\alpha;m_1}(\mathbf{k})| u_{\alpha;m_2}(\mathbf{k-q}}\braket{u_{\alpha;m_2}(\mathbf{k-q})|u_{\alpha,n_2}(\mathbf{k})}\notag\\
  = & \frac{1}{2}\sum_{\mathbf{q}} \sum_{\alpha \beta} V(\mathbf{q}) \rho_\alpha(\mathbf q)\rho_\beta(-\mathbf q)-\frac{1}{2} \sum_{\mathbf{q}} V(\mathbf{q}) \sum_{\mathbf{k}}\sum_{\alpha}\sum_{m} c^\dagger_{\alpha,m}(\mathbf{k})c_{\alpha,m}(\mathbf{k})
\end{align}
where we have used the completeness relation~\eqref{eq: completeness} for $\ket{u_{\alpha,m_2}(\mathbf{k-q})}$. Clearly the second term is just a chemical potential term.  For a fixed density, the normal ordering does not matter.

Next we project the Hamiltonian to the active bands. The Hilbert space is a tensor product ${\cal H}=\prod_m {\cal H}_m$, where ${\cal H}_m$ is the Hilbert space of band m. Suppose $M$ is the set of active bands and $\bar{M}$ is the set of all the other (fully filled or fully empty) bands. We have ${\cal H}={\cal H}_M\otimes {\cal H}_{\bar M}$. We want to project to the subspace ${\cal H}_S={\cal H}_M\otimes \ket{\Psi_{\bar M}}$, where $\ket{\Psi_{\bar M}}$ is a product state in $H_{\bar M}$; $\ket{\Psi_{\bar M}}$ is the Slater determinant such that each momentum is either occupied or empty depending on the specific band. 

We assume $\ket{\Psi_{\bar M}}$ is fixed and then we can trace it out to obtain an effective Hamiltonian in ${\cal H}_M$. The projection to ${\cal H}_S$ consists in calculating the expectation value of operators in ${\cal H}_{\bar M}$ for the state $\ket{\Psi_{\bar M}}$. The kinetic term is just a constant. We focus on the four fermion interaction in Eq.~\ref{eq:normal_order_interaction}.

Let us consider $c^\dagger_{\alpha,m_1}(\mathbf{k_1+q})c^\dagger_{\beta,n_1}(\mathbf{k_2-q})c_{\beta,n_2}(\mathbf{k_2})c_{\alpha,m_2}(\mathbf{k_1})$. We can group the terms using the number of indices belonging to $M$. If there is an odd number of indices belonging to $M$, the term vanishes in ${\cal H}_S$.
Therefore, we only need to consider two groups of terms: (1)$m_1,m_2,n_1,n_2 \in M$ (2)$m_1,n_2 \in M, m_2,n_1 \in \bar M$; (3)$m_1,m_2 \in M, n_1,n_2 \in \bar M$. (4) $n_1,m_2 \in M, m_1,n_2 \in \bar M$; (5)$n_1,n_2 \in M, m_1,m_2 \in \bar M$. Note that $m_1,n_1 \in M, m_2,n_2 \in \bar M$ vanishes.

Next we calculate the above five terms one by one.
The first term gives:
\begin{align}
   H_\mathrm{int}^{(1)}&=\frac{1}{2}\sum_{\mathbf{q}} \sum_{\alpha \beta} V(\mathbf q) :\tilde \rho_\alpha(\mathbf q)\tilde \rho_\beta(-\mathbf q):\notag\\
   &=\frac{1}{2}\sum_{\mathbf{q}} \sum_{\alpha \beta} V(\mathbf q) \tilde \rho_\alpha(\mathbf q)\tilde \rho_\beta(-\mathbf q)-\sum_{\alpha}\sum_{m,n \in M}\sum_{\mathbf k}\tilde\xi^\alpha_{mn}(\mathbf k)c^\dagger_{\alpha;m}(\mathbf k)c_{\alpha;n}(\mathbf k)
 \end{align} 
where 
\begin{equation}
\tilde \rho_\alpha(\mathbf q)=\sum_{m_1,m_2 \in M}\sum_{\mathbf k}c^\dagger_{\alpha,m_1}(\mathbf{k+q})c_{\alpha,m_2}(\mathbf{k})\braket{u_{\alpha,m_1}(\mathbf{k+q})| u_{\alpha,m_2}(\mathbf{k}}
\end{equation}
 \begin{align}
\tilde \xi^\alpha_{mn}(\mathbf k)&=\frac{1}{2}\sum_{\mathbf{q}}\sum_{m_2 \in M} V(\mathbf{q})\braket{u_{\alpha,m}(\mathbf{k})| u_{\alpha,m_2}(\mathbf{k-q}}\braket{u_{\alpha,m_2}(\mathbf{k-q})|u_{\alpha,n}(\mathbf{k})}
\label{eq: bilinear term 1}
 \end{align}
We find that if we do not use normal ordering, we need to add a bilinear term, which effectively behaves as a kinetic term proportional to the interaction strength.

The second term $H^{(2)}_\mathrm{int}$ is
\begin{align}
  H_\mathrm{int}^{(2)}&=-\frac{1}{2}\sum_{\mathbf{k_1},\mathbf{k_2},\mathbf{q}}\sum_{\alpha,\beta}\sum_{m_1,n_2\in M}\sum_{n_1,m_2\in \bar M} V(\mathbf{q})c^\dagger_{\alpha;m_1}(\mathbf{k_1+q})c_{\beta;n_2}(\mathbf{k_2})c^\dagger_{\beta;n_1}(\mathbf{k_2-q})c_{\alpha;m_2}(\mathbf{k_1})\notag\\
  &\braket{u_{\alpha;m_1}(\mathbf{k_1+q})| u_{\alpha;m_2}(\mathbf{k_1}}\braket{u_{\beta;n_1}(\mathbf{k_2-q})|u_{\beta;n_2}(\mathbf{k_2})}
\end{align}

Projecting to $H_S$ is equivalent to substituting $c^\dagger_{\beta;n_1}(\mathbf{k_2-q})c_{\alpha;m_2}(\mathbf{k_1})$ with its expectation values under $\ket{\Psi_{\bar M}}$. Defining $O\subset \bar M$ as the set of occupied bands. It is easy to find that
\begin{equation}
  H^{(2)}_\mathrm{int}=-\frac{1}{2}\sum_\alpha \sum_{\mathbf{k},\mathbf{q}}\sum_{m,n\in M}\sum_{m_2\in O} V(\mathbf{q})c^\dagger_{\alpha;m}(\mathbf{k})c_{\alpha;n}(\mathbf{k})\braket{u_{\alpha;m}(\mathbf{k})| u_{\alpha;m_2}(\mathbf{k-q}}\braket{u_{\alpha;m_2}(\mathbf{k-q})|u_{\alpha;n}(\mathbf{k})}
\end{equation}

The term (3) is
\begin{align}
  H_\mathrm{int}^{(3)}&=-\frac{1}{2}\sum_{\mathbf{k_1},\mathbf{k_2},\mathbf{q}}\sum_{\alpha,\beta}\sum_{m_1,m_2\in M}\sum_{n_1,n_2\in \bar M} V(\mathbf{q})c^\dagger_{\alpha;m_1}(\mathbf{k_1+q})c_{\beta;n_2}(\mathbf{k_2})c^\dagger_{\beta;n_1}(\mathbf{k_2-q})c_{\alpha;m_2}(\mathbf{k_1})\notag\\
  &\braket{u_{\alpha,m_1}(\mathbf{k_1+q})| u_{\alpha,m_2}(\mathbf{k_1}}\braket{u_{\beta,n_1}(\mathbf{k_2-q})|u_{\beta,n_2}(\mathbf{k_2})}
\end{align}
$\langle c^\dagger_{\beta;n_1}(\mathbf{k_2-q})c_{\beta;n_2}(\mathbf{k_2}) \rangle=0$ for $\mathbf{q}\neq 0$. As a result
\begin{align}
  H_\mathrm{int}^{(3)}&=-\frac{1}{2}\sum_{\mathbf{k}}\sum_{\alpha}\sum_{m,n\in M}4{\cal N}_O N_s V(0)c^\dagger_{\alpha,m}(\mathbf{k})  c_{\alpha,n}(\mathbf{k})
  \braket{u_{\alpha,m}(\mathbf{k})| u_{\alpha,n}(\mathbf{k}}
\end{align}
where ${\cal N}_O$ is the number of occupied bands and $N_s$ is the number of $k$ points, and $V(0) = U r_0$ for the screened Coulomb interaction Eq.~(10)
. This term is identical to the term (5). If there is only one active band ($m=n$), which is the case considered in this paper, then these terms are just a chemical potential which does not contribute to the bandwidth.

Term (4) is the same as the term (2) and gives
\begin{align}
  H_\mathrm{int}^{(4)}&=-\frac{1}{2}\sum_{\mathbf{k_1},\mathbf{k_2},\mathbf{q}}\sum_{\alpha,\beta}\sum_{m_1,n_2\in \bar M}\sum_{n_1,m_2\in M} V(\mathbf{q})c^\dagger_{\alpha;m_1}(\mathbf{k_1+q})c_{\beta;n_2}(\mathbf{k_2})c^\dagger_{\beta;n_1}(\mathbf{k_2-q})c_{\alpha;m_2}(\mathbf{k_1})\notag\\
  &\braket{u_{\alpha;m_1}(\mathbf{k_1+q})| u_{\alpha;m_2}(\mathbf{k_1}}\braket{u_{\beta;n_1}(\mathbf{k_2-q})|u_{\beta;n_2}(\mathbf{k_2})}
\end{align}
One can easily calculate the expectation value and get
\begin{align}
  H_\mathrm{int}^{(4)}=-\frac{1}{2}\sum_\alpha \sum_{\mathbf{k},\mathbf{q}}\sum_{m,n\in M}\sum_{m_2\in O} V(\mathbf{q})c^\dagger_{\alpha;m}(\mathbf{k})c_{\alpha;n}(\mathbf{k})\braket{u_{\alpha;m}(\mathbf{k})|u_{\alpha;m_2}(\mathbf{k+q})}\braket{u_{\alpha;m_2}(\mathbf{k+q})| u_{\alpha;n}(\mathbf{k}}
\end{align}
which is equal to $H_\mathrm{int}^{(2)}$.

Summing all five terms $H_\mathrm{int}^{(i)}$, we get the final form of projected interaction
\begin{equation}
  H_\mathrm{int}=\frac{1}{2}\sum_{\alpha \beta}\sum_{\mathbf{q}}V(\mathbf q):\tilde\rho_\alpha(\mathbf q)\rho_\beta(-\mathbf q):-\sum_{\alpha}\sum_{m,n \in M}\sum_{\mathbf k}\tilde{\tilde\xi}^\alpha_{mn}(\mathbf k)c^\dagger_{\alpha;m}(\mathbf k)c_{\alpha;n}(\mathbf k)
  \label{eq:final_interaction_normal_order}
\end{equation}

or equivalently

\begin{equation}
  H_\mathrm{int}=\frac{1}{2}\sum_{\alpha \beta}\sum_{\mathbf{q}}V(\mathbf q)\tilde \rho_\alpha(\mathbf q)\rho_\beta(-\mathbf q)-\sum_{\alpha}\sum_{m,n \in M}\sum_{\mathbf k}(\tilde \xi^\alpha_{mn}(\mathbf k)+\tilde{\tilde\xi}^\alpha_{mn}(\mathbf k))c^\dagger_{\alpha;m}(\mathbf k)c_{\alpha;n}(\mathbf k)
  \label{eq:final_interaction_normal_order}
\end{equation}

We have the additional bilinear terms $\tilde \xi^\alpha_{mn}(\mathbf k)$ (defined in Eq.~\eqref{eq: bilinear term 1}) and 
\begin{align}
\tilde{\tilde\xi}^\alpha_{mn}(\mathbf k)&=\sum_{\mathbf{q}}\sum_{m_2 \in O} V(\mathbf{q})\braket{u_{\alpha;m}(\mathbf{k})| u_{\alpha;m_2}(\mathbf{k-q}}\braket{u_{\alpha;m_2}(\mathbf{k-q})|u_{\alpha;n}(\mathbf{k})}\notag\\
& -4{\cal N}_O N_s V(0)
  \braket{u_{\alpha,m}(\mathbf{k})| u_{\alpha,n}(\mathbf{k}}
\end{align}

Finally, we have the full Hamiltonian in two equivalent forms.
\begin{equation}
  H_\mathrm{int}=\frac{1}{2}\sum_{\alpha \beta}\sum_{\mathbf{q}}V(\mathbf q):\tilde \rho_\alpha(\mathbf q)\rho_\beta(-\mathbf q):+H_K^1
  \label{eq:final_full_hamiltonian_normal_order}
\end{equation}
or
\begin{equation}
  H_\mathrm{int}=\frac{1}{2}\sum_{\alpha \beta}\sum_{\mathbf{q}}V(\mathbf q)\tilde \rho_\alpha(\mathbf q)\rho_\beta(-\mathbf q):+H_K^2
  \label{eq:final_full_hamiltonian_no_normal_order}
\end{equation}

where we denote $H_K^0$ as the bare dispersion, and the effective kinetic terms are
\begin{equation}
  H_K^1=H_K^0-\sum_{\alpha}\sum_{m,n \in M}\sum_{\mathbf k}\tilde{\tilde\xi}^\alpha_{mn}(\mathbf k)c^\dagger_{\alpha;m}(\mathbf k)c_{\alpha;n}(\mathbf k)
\end{equation}
and
\begin{equation}
  H_K^2=H_K^0-\sum_{\alpha}\sum_{m,n \in M}\sum_{\mathbf k}\left(\tilde \xi^\alpha_{mn}(\mathbf k)+\tilde{\tilde\xi}^\alpha_{mn}(\mathbf k)\right)c^\dagger_{\alpha;m}(\mathbf k)c_{\alpha;n}(\mathbf k)
\end{equation}
In $H^1_K$, we add Hartree-Fock terms (actually only the Fock term) originating from occupied bands. For the graphene moir\'e terms, we expect that the bare band structure from the continuum model will be greatly renormalized by this effect. In $H^2_K$, we further include a dispersion term from the active band itself.

Using $H^0_K$, $H^1_K$ and $H^2_K$ we can define three bandwidths $W_\mathrm{bare}$ (the bare bandwidth), $W_1$ and $W_2$.
In the main text, we have used $W_2 = 0$ for our analytical argument; in this limit, we have shown that the spin and valley polarized state is an exact ground state of the interaction. In the numerical exact diagonalizations, we have used $W_1=0$ as the flat-band limit; in this case there is no guarantee that the spin-valley polarized state is a ground state, and in fact, our results show examples where it is an excited state.

In a different perspective, we can calculate the effective dispersion at a certain filling $\nu_T$. In a quantum Hall ferromagnetic state, we can assume a flavor $\alpha$ is fully filled. The corresponding dispersion from Hartree-Fock calculation for the band $\alpha$ is:
\begin{equation}
  H_K^\mathrm{eff}=H_K^1-\sum_{m,n \in M}\sum_{\mathbf k}\left(2\tilde \xi^\alpha_{mn}(\mathbf k)+\tilde{\tilde\xi}^\alpha_{mn}(\mathbf k)\right)c^\dagger_{\alpha;m}(\mathbf k)c_{\alpha;n}(\mathbf k)
\end{equation}
We can again define a bandwidth $W_\mathrm{eff}$ for the fully filled state. Note that there is an additional factor of 2. Therefore for the active band $W_\mathrm{eff}$ is again different from $W_2$.  Compared to $W_1$, $W_\mathrm{eff}$ includes the Hatree-Fock term from the active band itself. However, $W_2$ only includes one half of this term. In that sense, it is surprising that $W_2=0$ is the condition for flat band ferromagnetism.

\section{Additional numerical results}

\begin{figure}[ht]
    \centering
        \includegraphics[width=0.49 \textwidth]{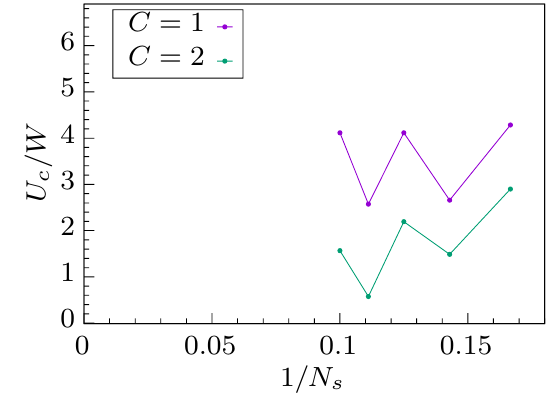}
        \caption{Onset of ferromagnetism in the valley-polarized model $H_1$ with Chern number $C = 1, 2$ (the exact parameters are set in the main text). The results are the same for the spin-polarized model within our numerical precision. In finite-size, the ground state within the $S_z=0$ sector is especially stable at $U<U_c$, resulting in an even/odd effect in the value of $U_c$.}
    \label{fig:ED 2 band}
\end{figure}

In this section, we give additional numerical results on the two-band (Honeycomb) model
$H_1$ Eq.~(13) of the main text.

\subsection{Exact diagonalization results}
First, we complement our DMRG study with results of exact diagonalization (ED). We limit ourselves to the spin-polarized or valley-polarized versions of $H_1$, since only very few system sizes are accessible within the full (non-polarized) model using ED. Our results are summarized in Fig.~\ref{fig:ED 2 band}. We find that the onset of ferromagnetism is the same in these two limits in finite size, and thus show only one plot. The results are consistent with the results of our DMRG study, with small differences which can be accounted for by invoking finite-size effects.

\subsection{Bond dimension convergence of the DMRG results}
Due to the Wannier obstruction of topological bands, it is necessary for a real space simulation to include both conduction and valence bands. This is highly challenging even in the parameter regimes where insulating ground states are expected, because a fully filled Chern band still carries large entanglement. More specifically, even though the insulating (fully-polarized) ground state is a product state in the (momentum-space) basis of Bloch wave functions, the entanglement entropy associated with a real-space cut is large. As a result, the critical value of the interaction $U_c$ still displays relatively large variations with bond dimension in our results (see Fig.~3c of the main text).

\begin{figure}[ht]
    \centering
        \includegraphics[width=0.45 \textwidth]{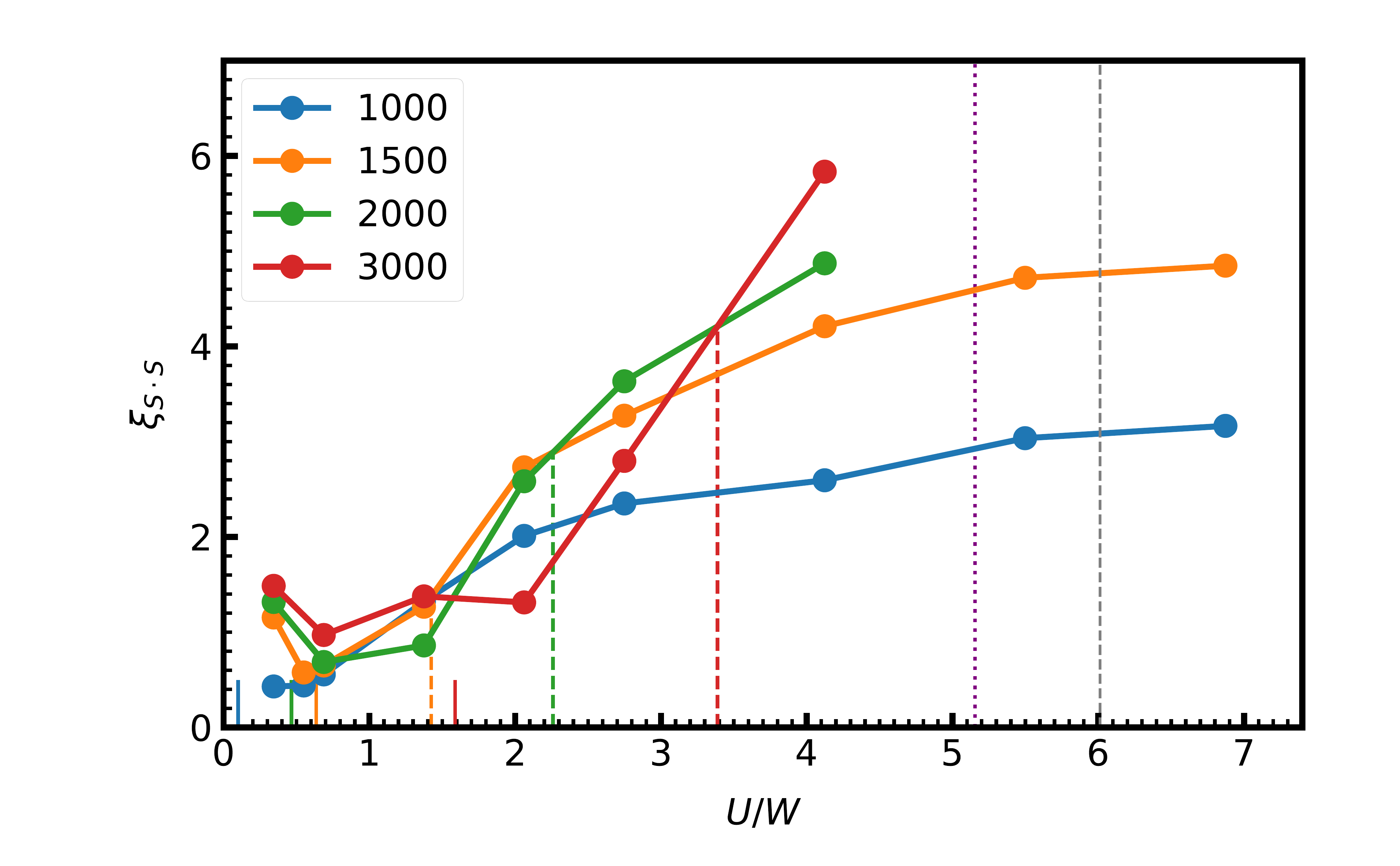}
        \caption{Correlation length $\xi_{S\cdot S}$ associated with the $\langle S\cdot S \rangle $ correlations for the $C=1$ model at $\nu_T=1$ as extracted from the $S_z=0$, $I_z=1$ ground state. The various colors correspond to different values of the bond dimension. The dashed vertical lines indicate the onset of ferromagnetism $U_c/W$ based on the correlation length; beyond the line, $\xi_{S\cdot S}$ increases when the bond dimension increases. The solid lines indicate the onset of ferromagnetism based on the energies shown in Fig.~3 of the main text. Even here, $U_c/W$ is still smaller than the band gap (gray dashed line) and seemingly approaches the value found in the valley-polarized model (purple dashed line).}
    \label{fig:quarter E-U}
\end{figure}

In the main text we have relied on energetics to detect the onset of ferromagnetism; we ran independent DMRG simulations for each spin and valley polarization sector, and compared all these energies to infer the spin and valley polarization of the ground state. Alternatively, we can use the fact that the correlation length associated with spin-spin correlations $\xi_{S\cdot S}$ diverges in the ferromagnetic state. Since $\xi_{S\cdot S}$ is very sensitive to the bond dimension, this method provides a more conservative estimate for $U_c/W$. The evolution of $\xi_{S\cdot S}$ with the interaction strength is shown in Fig.~\ref{fig:quarter E-U} for various values of the bond dimension. It shows us that finite bond dimension effects are still important at the bond dimensions we used. Nevertheless, even with this more conservative method, $U_c/W$ is still significantly smaller than the band gap.

\end{document}